\documentclass[12pt]{iopart}
\usepackage{amssymb}
\usepackage{euscript}
\usepackage[english]{babel}

\newcommand{\eqref}[1]{(\ref{#1})}

\begin{document}

\title[Lax representations and exotic cohomology of symmetry algebras]{%
Lax representations with non-removable parameter and exotic cohomology
of symmetry algebras of PDEs}

\author[O.I. Morozov]{Oleg I. Morozov}

\vskip 10 pt

\address{
  Faculty of Applied
  Mathematics, AGH University of Science and Technology,
  \\
  Al. Mickiewicza 30,
  Cracow 30-059, Poland;
    \\
  ~~
  \\
  Institute of Control Sciences of Russian Academy of Sciences,
  \\
  Profsoyuznaya 65, Moscow 117997, Russia;
  \\
  ~
  \\
  e-mail: morozov\symbol{64}agh.edu.pl
}



\begin{abstract}
This paper develops the technique of constructing Lax representations for {\sc pde}s via non-central extensions
of their contact symmetry algebras.  We show that the me\-thod is applicable to the Lax representations with
non-removable spectral pa\-ra\-me\-ters.
\end{abstract}

\maketitle


\section{Introduction}

Lax representations, also known as zero-curvature representations, Wahlquist--Es\-ta\-brook prolongation
structures, inverse scattering transformations, or differential co\-ve\-rings
\cite{KrasilshchikVinogradov1984,KrasilshchikVinogradov1989}, are a key feature of integrable partial
differential equations ({\sc pde}s) and a starting set\-ting for a number of techniques of studying them
such as B\"acklund trans\-for\-ma\-ti\-ons, Dar\-boux transformations, recursion operators, nonlocal symmetries,
and nonlocal con\-ser\-va\-tion laws. Lax representations with non-removable (spectral) pa\-ra\-me\-ter are of
special interest in the theory of integrable {\sc pde}s, see, e.g.,
\cite{AblowitzClarkson1991,Das1989,DoddFordy1983,TakhtadzhyanFaddeev1987}.
The chal\-len\-ging unsolved pro\-blem in this theory is to find conditions that are formulated in inherent terms
of a {\sc pde} under study  and ensure existence of a Lax representation for the {\sc pde}. Recently, an approach
to this problem has been proposed in \cite{Morozov2017,Morozov2018a}, where it was shown that for some
{\sc pde}s their Lax representations can be inferred from the second exotic cohomology of the contact symmetry
algebras of the {\sc pde}s.

The present paper provides an important supplement to the technique of \cite{Morozov2017,Morozov2018a}.
Na\-me\-ly, we show that Lax representations with a non-removable parameter arise na\-tu\-ral\-ly from non-central
extensions of the symmetry algebras generated by nontrivial se\-cond exotic cohomology groups. We consider here
two equations: the hyper-CR Ein\-stein--Weyl equation \cite{Kuzmina1967,Mikhalev1992,Pavlov2003,Dunajski2004}
\begin{equation}
u_{yy} = u_{tx} + u_y\,u_{xx} - u_x\,u_{xy}
\label{Pavlov_eq}
\end{equation}
with the Lax representation
\begin{equation}
\left\{
\begin{array}{lcl}
v_t &=& (\lambda^2+\lambda\,u_x-u_y)\,v_x,
\\
v_y &=& -(u_x+\lambda)\,v_x,
\end{array}
\right.
\label{Pavlov_eq_covering}
\end{equation}
and the four-dimensional  equation
\begin{equation}\label{4D_UHE}
u_{zz} = u_{tx}+u_x u_{yz}- u_z u_{xy}
\end{equation}
which was introduced in \cite{BogdanovPavlov2016}. The 3-dimensional reduction of \eqref{4D_UHE} defined by
substitution for $u_t=0$ produces the universal hierarchy equation
\cite{MartinezAlonsoShabat2002,MartinezAlonsoShabat2004}
\begin{equation*}
u_{zz} = u_x u_{yz}- u_z u_{xy},
\end{equation*}
therefore we refer equation \eqref{4D_UHE} as the {\it four-dimensional universal hierarchy equation}.
The Lax representation
\begin{equation}\label{Pavlov_Stoilov_covering}
\left\{
\begin{array}{lcl}
v_t&=& \lambda^2\,v_x +(\lambda\,u_x+u_z)\,v_y,
\\
v_z&=& \lambda\,v_x+u_x\,v_y
\end{array}
\right.
\end{equation}
for \eqref{4D_UHE} was found in \cite{PavlovStoilov2017}.

The parameters $\lambda$ in  \eqref{Pavlov_eq_covering} and \eqref{Pavlov_Stoilov_covering}
are non-removable. This assertion can be proven  by the
method of \cite[Sections 3.2, 3.6]{KrasilshchikVinogradov1989},
\cite{Krasilshchik2000,IgoninKrasilshchik2000,Marvan2002,Cieslinski1992,Cieslinski1993a,Cieslinski1993b},
see Remarks 1 and 3 be\-low.

The following structure distinguishes the contact symmetry algebras for both equations: they are semi-direct
products $\mathfrak{s}_\infty \rtimes \mathfrak{s}_\diamond$ of an (invariantly defined) infinite-di\-men\-si\-o\-nal
ideal $\mathfrak{s}_\infty$ and a non-Abelian  finite-dimensional Lie algebra $\mathfrak{s}_\diamond$. The
se\-cond exotic cohomology groups of the finite-dimensional subalgebras $\mathfrak{s}_\diamond$ appear to be
non\-tri\-vi\-al for both equations, and the corresponding nontrivial 2-cocycles produce non-central ex\-ten\-si\-ons of
their symmetry algebras $\mathfrak{s}_\infty \rtimes \mathfrak{s}_\diamond$. We show that certain linear
combination of the Maurer--Cartan forms of the ex\-ten\-si\-on of the symmetry algebra defines the Lax
re\-pre\-sen\-ta\-ti\-on \eqref{Pavlov_eq_covering} for equation \eqref{Pavlov_eq}. For equa\-ti\-on
\eqref{4D_UHE} we apply the extension procedure twice and show that the Maurer--Cartan forms of the second
ex\-ten\-si\-on produce the Lax re\-pre\-sen\-ta\-ti\-on \eqref{Pavlov_Stoilov_covering}.

\section{Preliminaries}

All considerations in this paper are local. All functions are assumed to be real-analytic.

\subsection{Symmetries and differential coverings}
The presentation in this subsection closely follows
\cite{KrasilshchikVerbovetsky2011,KrasilshchikVerbovetskyVitolo2012},
see also \cite{KrasilshchikVinogradov1984,KrasilshchikVinogradov1989,VK1999}.
Let $\pi \colon \mathbb{R}^n \times \mathbb{R}^m \rightarrow \mathbb{R}^n$,
$\pi \colon (x^1, \dots, x^n, u^1, \dots, u^m) \mapsto (x^1, \dots, x^n)$, be a trivial bundle, and
$J^\infty(\pi)$ be the bundle of its jets of the infinite order. The local coordinates on $J^\infty(\pi)$ are
$(x^i,u^\alpha,u^\alpha_I)$, where $I=(i_1, \dots, i_n)$ are multi-indices, and for every local section
$f \colon \mathbb{R}^n \rightarrow \mathbb{R}^n \times \mathbb{R}^m$ of $\pi$ the corresponding infinite jet
$j_\infty(f)$ is a section $j_\infty(f) \colon \mathbb{R}^n \rightarrow J^\infty(\pi)$ such that
$u^\alpha_I(j_\infty(f))
=\displaystyle{\frac{\partial ^{\#I} f^\alpha}{\partial x^I}}
=\displaystyle{\frac{\partial ^{i_1+\dots+i_n} f^\alpha}{(\partial x^1)^{i_1}\dots (\partial x^n)^{i_n}}}$.
We put $u^\alpha = u^\alpha_{(0,\dots,0)}$. Also, we will simplify notation in the following way, e.g., in the
case of $n=4$, $m=1$: we denote $x^1 = t$, $x^2= x$, $x^3= y$, $x^4=z$ and
$u^1_{(i,j,k,l)}=u_{{t \dots t}{x \dots x}{y \dots y}{z \dots z}}$ with $i$  times $t$,
$j$  times $x$, $k$  times $y$, and $l$ times $z$.

The  vector fields
\begin{equation*}
D_{x^k} = \frac{\partial}{\partial x^k} + \sum \limits_{\# I \ge 0} \sum \limits_{\alpha = 1}^m
u^\alpha_{I+1_{k}}\,\frac{\partial}{\partial u^\alpha_I},
\qquad k \in \{1,\dots,n\},
\end{equation*}
$(i_1,\dots, i_k,\dots, i_n)+1_k = (i_1,\dots, i_k+1,\dots, i_n)$,  are called {\it total derivatives}.
They com\-mu\-te everywhere on
$J^\infty(\pi)$:  $[D_{x^i}, D_{x^j}] = 0$.

The {\it evolutionary vector field} associated to an arbitrary vector-valued smooth function
$\varphi \colon J^\infty(\pi) \rightarrow \mathbb{R}^m $ is the vector field
\[
\mathbf{E}_{\varphi} = \sum \limits_{\# I \ge 0} \sum \limits_{\alpha = 1}^m
D_I(\varphi^\alpha)\,\frac{\partial}{\partial u^\alpha_I}
\]
with $D_I=D_{(i_1,\dots\,i_n)} =D^{i_1}_{x^1} \circ \dots \circ D^{i_n}_{x^n}$.

A system of {\sc pde}s $F_r(x^i,u^\alpha_I) = 0$ of the order $s \ge 1$ with $\# I \le s$,
$r \in \{1,\dots, R\}$ for some $R \ge 1$,
defines the submanifold
$\EuScript{E}=\{(x^i,u^\alpha_I)\in J^\infty(\pi)\,\,\vert\,\,D_K(F_r(x^i,u^\alpha_I))=0,\,\,\# K\ge 0\}$
in $J^\infty(\pi)$.

A function $\varphi \colon J^\infty(\pi) \rightarrow \mathbb{R}^m$ is called a {\it (generator of an
infinitesimal) symmetry} of equation $\EuScript{E}$ when $\mathbf{E}_{\varphi}(F) = 0$ on $\EuScript{E}$. The
symmetry $\varphi$ is a solution to the {\it defining system}
\begin{equation}
\ell_{\EuScript{E}}(\varphi) = 0,
\label{defining_eqns}
\end{equation}
where $\ell_{\EuScript{E}} = \ell_F \vert_{\EuScript{E}}$ with the matrix differential operator
\begin{equation*}
\ell_F = \left(\sum \limits_{\# I \ge 0}\frac{\partial F_r}{\partial u^\alpha_I}\,D_I\right).
\end{equation*}
The {\it symmetry algebra} $\mathfrak{sym} (\EuScript{E})$ of equation $\EuScript{E}$ is the linear space of
solutions to  (\ref{defining_eqns}) endowed with the structure of a Lie algebra over $\mathbb{R}$ by the
{\it Jacobi bracket} $\{\varphi,\psi\} = \mathbf{E}_{\varphi}(\psi) - \mathbf{E}_{\psi}(\varphi)$.
The {\it algebra of contact symmetries} $\mathfrak{sym}_0 (\EuScript{E})$ is the Lie subalgebra of $\mathfrak{sym} (\EuScript{E})$
defined as $\mathfrak{sym} (\EuScript{E}) \cap J^1(\pi)$.

Consider $\EuScript{W} = \mathbb{R}^\infty$ with  coordinates $w^s$, $s \in  \mathbb{N} \cup \{0\}$. Locally,
an (infinite-di\-men\-si\-o\-nal)  {\it differential covering} of $\EuScript{E}$ is a trivial bundle
$\tau \colon J^\infty(\pi) \times \EuScript{W} \rightarrow J^\infty(\pi)$
equipped with {\it extended total derivatives}
\[
\widetilde{D}_{x^k} = D_{x^k} + \sum \limits_{ s =0}^\infty
T^s_k(x^i,u^\alpha_I,w^j)\,\frac{\partial }{\partial w^s}
\]
such that $[\widetilde{D}_{x^i}, \widetilde{D}_{x^j}]=0$ for all $i \not = j$ whenever
$(x^i,u^\alpha_I) \in \EuScript{E}$. Define
the partial derivatives of $w^s$ by  $w^s_{x^k} =  \widetilde{D}_{x^k}(w^s)$.  This yields the system of
{\it covering equations}
\begin{equation}
w^s_{x^k} = T^s_k(x^i,u^\alpha_I,w^j),
\label{WE_prolongation_eqns}
\end{equation}
which is compatible whenever $(x^i,u^\alpha_I) \in \EuScript{E}$.

Dually, the covering is defined by the {\it Wahlquist--Estabrook forms}
\begin{equation}
d w^s - \sum \limits_{k=1}^{m} T^s_k(x^i,u^\alpha_I,w^j)\,dx^k
\label{WEfs}
\end{equation}
as follows: when $w^s$  and $u^\alpha$ are considered to be functions of $x^1$, ... , $x^n$, forms \eqref{WEfs} are equal
to zero whenever system \eqref{WE_prolongation_eqns} holds.

\subsection{Exotic cohomology of Lie algebras}

For a Lie algebra
 $\mathfrak{g}$ over $\mathbb{R}$, its representation $\rho \colon \mathfrak{g} \rightarrow \mathrm{End}(V)$,
and $k \ge 1$
let $C^k(\mathfrak{g}, V) =\mathrm{Hom}(\Lambda^k(\mathfrak{g}), V)$
be the space of all $k$--linear skew-symmetric mappings from $\mathfrak{g}$ to $V$. Then
the Chevalley--Eilenberg differential
complex
\begin{equation*}
V=C^0(\mathfrak{g}, V) \stackrel{d}{\longrightarrow} C^1(\mathfrak{g}, V)
\stackrel{d}{\longrightarrow} \dots \stackrel{d}{\longrightarrow}
C^k(\mathfrak{g}, V) \stackrel{d}{\longrightarrow} C^{k+1}(\mathfrak{g}, V)
\stackrel{d}{\longrightarrow} \dots
\end{equation*}
is generated by the differential $d \colon \theta \mapsto d\theta$ such that
\begin{equation*}
d \theta (X_1, ... , X_{k+1}) =
\sum\limits_{q=1}^{k+1}
(-1)^{q+1} \rho (X_q)\,(\theta (X_1, ... ,\hat{X}_q, ... ,  X_{k+1}))
\end{equation*}
\begin{equation*}
\quad
+\sum\limits_{1\le p < q \le k+1} (-1)^{p+q}
\theta ([X_p,X_q],X_1, ... ,\hat{X}_p, ... ,\hat{X}_q, ... ,  X_{k+1}).
\end{equation*}
The cohomology groups of the complex $(C^{*}(\mathfrak{g}, V), d)$ are referred to as
the {\it cohomology groups of the Lie algebra} $\mathfrak{g}$ {\it with coefficients in the representation}
$\rho$. For the trivial representation $\rho_0 \colon \mathfrak{g} \rightarrow \mathbb{R}$,
$\rho_0 \colon X \mapsto 0$, the cohomology groups are denoted by
$H^{*}(\mathfrak{g})$.

Consider a Lie algebra $\mathfrak{g}$ over $\mathbb{R}$ with non-trivial first cohomology group
$H^1(\mathfrak{g})$ and take a closed 1-form $\alpha$ on $\mathfrak{g}$ such that $[\alpha] \neq 0$.
Then for any $c \in \mathbb{R}$
define new differential
$d_{c \alpha} \colon C^k(\mathfrak{g},\mathbb{R}) \rightarrow C^{k+1}(\mathfrak{g},\mathbb{R})$ by
the formula
\begin{equation*}
d_{c \alpha} \theta = d \theta - c \,\alpha \wedge \theta.
\end{equation*}
From  $d\alpha = 0$ it follows that
$d_{c \alpha} ^2=0$. The cohomology groups of the complex
\begin{equation*}
C^1(\mathfrak{g}, \mathbb{R})
\stackrel{d_{c \alpha}}{\longrightarrow}
\dots
\stackrel{d_{c \alpha}}{\longrightarrow}
C^k(\mathfrak{g}, \mathbb{R})
\stackrel{d_{c \alpha}}{\longrightarrow}
C^{k+1}(\mathfrak{g}, \mathbb{R})
\stackrel{d_{c \alpha}}{\longrightarrow} \dots
\end{equation*}
are referred to as the {\it exotic} {\it cohomology groups}  \cite{Novikov2002,Novikov2005} of $\mathfrak{g}$ and denoted by
$H^{*}_{c\alpha}(\mathfrak{g})$.

\section{Hyper-CR Einstein--Weyl equation}

\subsection{Contact symmetries}

Direct computations\footnote[1]{We carried out computations of generators of contact symmetries and their
commutator tables in the {\it Jets} software \cite{Jets}.} show that the Lie algebra of the contact symmetries
$\mathfrak{sym}_0(\EuScript{E}_1)$ of the hyper-CR Einstein--Weyl equation $\EuScript{E}_1$ is generated by the
functions
\[
\phi_0(A) = -A\,u_t -\left(x\,A^{\prime} +\case{1}{2}\,y^2\,A^{\prime\prime}\right)\,u_x
-y\,A^{\prime}\, u_y+u\,A^{\prime} +x\,y\,A^{\prime\prime} + \case{1}{6}\,y^3\,A^{\prime\prime\prime},
\]
\[
\phi_1(A) = - y\,A^{\prime}\,u_x- A\,u_y+x\,A^{\prime} + \case{1}{2}\,y^2\,A^{\prime\prime},
\]
\[
\phi_2(A) = -A\,u_x+y\,A^{\prime},
\]
\[
\phi_3(A) = A,
\]
\[
\psi_0 = -2\,x\,u_x-y\,u_y+3\,u,
\]
\[
\psi_1 = - y\,u_x+2\,x
\]
where $A=A(t)$ and $B=B(t)$ below are arbitrary functions of $t$. The commutators of the generators are given
by equations
\begin{equation}
\left\{
\begin{array}{lcl}
\{\phi_i(A), \phi_j(B)\}  &=&  \phi_{i+j}(A\,B^{\prime} - B\,A^{\prime}),
\\
\{\psi_i, \phi_k(A)\} &=& -k\,\phi_{k+i}(A),
\\
\{\psi_0, \psi_1\} &=& -\psi_1,
\end{array}
\right.
\label{Pavlov_eq_commutator_table}
\end{equation}
where $\phi_k(A)=0$ for $k > 3$.
From equations \eqref{Pavlov_eq_commutator_table} it follows that the contact symmetry algebra of equation
\eqref{Pavlov_eq} is the semi-direct product
$\mathfrak{sym}_0(\EuScript{E}_1) =\mathfrak{b}_{\infty} \rtimes \mathfrak{b}_\diamond$
of the two-dimensional non-Abelian Lie algebra $\mathfrak{b}_\diamond =\langle \psi_0, \psi_1 \rangle$
and the infinite-dimensional
ideal\footnote{Here and below we use notation
$\EuScript{D}(\mathfrak{g})=\EuScript{D}^1(\mathfrak{g}) = [\mathfrak{g}, \mathfrak{g}]$,
$\EuScript{D}^{k+1}(\mathfrak{g}) = [\EuScript{D}^{k}(\mathfrak{g}),\EuScript{D}^{k}(\mathfrak{g})]$
for the derived se\-ri\-es of a Lie algebra $\mathfrak{g}$.}
$\mathfrak{b}_{\infty} = \EuScript{D}(\mathfrak{sym}_0(\EuScript{E}_1))=
\langle \phi_k(A) \,\,\,\vert\,\,\, 0\le k \le 3 \rangle$, which, in its turn, is isomorphic to the tensor
product $\mathbb{R}_3[h] \otimes \mathfrak{w}$ of the (commutative associative) algebra of truncated polynomials
$\mathbb{R}_3[h] = \mathbb{R}[h]/\langle h^4 = 0\rangle$ and the Lie algebra
$\mathfrak{w} = \langle t^n \partial_t \,\,\,\vert\,\,\, n \in \mathbb{Z}_{+} \rangle$.

\vskip 5 pt
\noindent
{\sc Remark 1}.
The spectral parameter $\lambda$ in the covering \eqref{Pavlov_eq_covering} is non-removable, that is, it cannot
be eliminated from \eqref{Pavlov_eq_covering} by a trans\-for\-ma\-ti\-on of the covering that is identical on
the base equation $\EuScript{E}_1$. In accordance with
\cite{KrasilshchikVinogradov1989,Krasilshchik2000,IgoninKrasilshchik2000}, to prove this claim it
suffices to note that it is impossible to lift the symmetry $\psi_1$ of equation \eqref{Pavlov_eq}  to a symmetry
of system \eqref{Pavlov_eq_covering}. Then for the  vector field
$V(\psi_1) = y\,\partial_x + 2\,x\,\partial_u$ associated to the ge\-ne\-ra\-tor $\psi_1$ and
the Wahlquist--Estabrook form
$\omega =  dv+u_y\,v_x\,dt -v_x\,dx +u_x\, v_x\,dy$ that defines covering \eqref{Pavlov_eq_covering} with
$\lambda = 0$ we have
$e^{-\lambda V(\psi_1)} \omega= dv-(\lambda^2+\lambda\,u_x-u_y)\,v_x\,dt-v_x\,dx+(u_x+\lambda)\,v_x\,dy$.
This is the Wahlquist--Estabrook form of covering \eqref{Pavlov_eq_covering} with $\lambda \in \mathbb{R}$.
\hfill $\diamond$
\vskip 5 pt

Consider the basis of $\mathfrak{sym}_0(\EuScript{E}_1)$ given by
generators $\psi_0$, $\psi_1$, and $\phi_k(t^n)$ with $k \in \{0, \dots,  3\}$, $n \in \mathbb{Z}_{+}$.
Define the Maurer--Cartan forms $\alpha_0$, $\alpha_1$, $\theta_{k,n}$  for $\mathfrak{sym}_0(\EuScript{E}_1)$
as the dual forms to this basis:
$\alpha_i(\psi_j) = \delta_{ij}$,
$\alpha_i(\phi_k(t^n)) = 0$,
$\theta_{k,n}(\psi_i) = 0$,
$\theta_{k,n}(\phi_l(t^m)) = \delta_{kl} \delta_{nm}$.
Put
$\Theta = \sum \limits_{k=0}^{3} \sum \limits_{m=0}^{\infty} \frac{h_1^m}{m!}\,h_2^k\,\theta_{k,m}$,
where $h_1$ and $h_2$ are (formal) parameters such that $dh_i=0$, and denote $\nabla_i =\partial_{h_i}$. Then the
commutator table \eqref{Pavlov_eq_commutator_table} gives the {\it Maurer--Cartan structure equations}
\begin{equation}
\left\{
\begin{array}{lcl}
d\alpha_0 &=& 0,
\\
d\alpha_1 &=& \alpha_0 \wedge \alpha_1,
\\
d \Theta &=& \nabla_1 (\Theta) \wedge \Theta + (h_2\,\alpha_0 + h_2^2\,\alpha_1) \wedge \nabla_2 (\Theta)
\end{array}
\right.
\label{structure_equations_for_Pavlov_eq}
\end{equation}
of $\mathfrak{sym}_0(\EuScript{E}_1)$.

\subsection{Second exotic cohomology group and non-central extension}

From the structure equations \eqref{structure_equations_for_Pavlov_eq} it follows that
$H^1(\mathfrak{sym}_0(\EuScript{E}_1)) = \mathbb{R} \alpha_0$
and
\[
H^2_{c\,\alpha_0}(\mathfrak{b}_\diamond)  =
\left\{
\begin{array}{lcl}
\langle [\alpha_0 \wedge \alpha_1]\rangle, &~~~& c =1,
\\
\{[0]\}, && c \neq 1.
\end{array}
\right.
\]
Moreover, we have
$H^2_{\alpha_0}(\mathfrak{b}_\diamond) \subseteq H^2_{\alpha_0}(\mathfrak{sym}_0(\EuScript{E}_1))$. Hence the
nontrivial 2-cocycle $\alpha_0 \wedge \alpha_1$ of the differential $d_{\alpha_0}$ defines a non-central
extension $\widehat{\mathfrak{b}}_\diamond$ of the Lie algebra $\mathfrak{b}_\diamond$ and thus a non-central
extension $\mathfrak{b}_\infty \rtimes \widehat{\mathfrak{b}}_\diamond$ of the Lie algbera
$\mathfrak{sym}_0(\EuScript{E}_1)$. The additional Maurer--Cartan form $\sigma$ for the extended Lie algebra is
a solution to $d_{\alpha_0} \sigma = \alpha_0 \wedge \alpha_1$, that is, to equation
\begin{equation}
d \sigma = \alpha_0 \wedge \sigma + \alpha_0 \wedge \alpha_1.
\label{extension_se_1}
\end{equation}
This equation is automatically compatible with the structure equations \eqref{structure_equations_for_Pavlov_eq}
of the Lie algebra $\mathfrak{sym}_0(\EuScript{E}_1)$.

\subsection{Maurer--Cartan forms and Lax representation}

For the purposes of the present paper we need explicit expressions for the Maurer--Cartan forms
$\alpha_i$, $\theta_{k,0}$, and $\sigma$.  We can compute them  via two approaches. The first one is to
integrate equations \eqref{structure_equations_for_Pavlov_eq}, \eqref{extension_se_1} step by step. Each
integration gives certain number of  new coordinates (the `integration constants') to express the new
form, while it is not clear how these coordinates relate to the coordinates of $\EuScript{E}_1$.
For example, integrating the first two equations from system
\eqref{structure_equations_for_Pavlov_eq} and then equation \eqref{extension_se_1} we obtain
\[
\alpha_0 = \frac{dq}{q},
\quad
\alpha_1 = q\, ds,
\quad
\sigma  = q\,(dv+\ln q \,ds),
\]
where $q$, $s$, and $v$ are free parameters and $q>0$ (we put $\alpha_0 = dq/q$ instead of the natural choice
$\alpha_0 = dq$ to simplify the further computations).
The second approach to computing  the Maurer--Cartan forms is to use Cartan's method of equivalence,
\cite{Cartan1,Cartan2,Cartan3,Cartan4,Olver1995,FelsOlver1998},
see details and examples of applying the method to symmetry pseudo-groups to {\sc pde}s in
\cite{Morozov2002,Morozov2006,Morozov2009b}.
In case of the structure equations \eqref{structure_equations_for_Pavlov_eq} this technique
shows that
\begin{enumerate}
\item[{\it(i)}] $\theta_{0,0}$ is a multiple of $dt$, $\theta_{1,0}$ is a linear combination of $dy$, $dt$,
$\theta_{2,0}$ is a linear com\-bi\-na\-ti\-on of $dx$, $dy$, $dt$,
\item[{\it(ii)}] $\theta_{3,0}$ is a multiple
of the contact form $du -u_t\,dt - u_x\,dx - u_y\,dy$.
\end{enumerate}
Using {\it(i)} we have
$\theta_{0,0} = a_0\,dt$,
$\theta_{1,0} = a_0\,q\,(dy +a_1\,dt)$,
$\theta_{2,0} = a_0\,q^2\,(dx + (a_1+s)\,dy + a_2\,dt)$,
with new parameters $a_0 \neq 0$, $a_1$, $a_2$, while
{\it (ii)} then gives $a_1 = - u_x-2\,s$, $a_2  =- u_y +s\,u_x+s^2$, and
$\theta_{3,0} = a_0\,q^3\,(du -u_t\,dt - u_x\,dx - u_y\,dy)$.

Consider the linear combination
\[
\sigma - \theta_{2,0} =q\,\left(dv + \ln q \,ds -a_0q\,\left(dx -(s+u_x)\,dy+(s^2+s\,u_x-u_y)\,dt\right)\right)
\]
and assume that $u$ and $v$ are functions of $t$, $x$, $y$. Then $\sigma - \theta_{2,0}=0$ implies
$q = \exp (-v_s)$,
$a_0 = v_x\,\exp (v_s)$.
After this change of notation we obtain the Wahlquist--Estabrook form
\[
\sigma - \theta_{2,0} =\mathrm{e}^{-v_s}\,\left(dv -v_s\,ds -v_x\,dx +(s+u_x)\,dy-(s^2+s\,u_x-u_y)\,dt\right)
\]
of the covering
\[
\left\{
\begin{array}{lcl}
v_t &=& (s^2+s\,u_x-u_y)\,v_x,
\\
v_y &=& -(u_x+s)\,v_x.
\end{array}
\right.
\]
This system differs from \eqref{Pavlov_eq_covering} by notation.

\vskip 5 pt
\noindent
{\sc Remark 2}.
The non-removable parameter $s$ in the above system has appeared during com\-pu\-tation of  form $\alpha_1$,
 which is dual to the unliftable symmetry $\psi_1$, cf. Remark 1.
\hfill $\diamond$

\section{The four-dimensional universal hierarchy equation}

\subsection{Contact symmetries}

The generators of the Lie algebra $\mathfrak{sym}_0(\EuScript{E}_2)$ of the contact symmetries of equation
\eqref{4D_UHE} are
\begin{equation*}
\begin{array}{rcl}
\phi_0(A) &=& -A\,u_y+ A_y\,u-A_t\,z,
\\
\phi_1(A) &=& A,
\\
\psi_1 &=& -t\,u_t+x\,u_x-u,
\\
\psi_2 &=& -u_t,
\\
\psi_3 &=& -2\,x\,u_x-z\,u_z+u,
\\
\psi_4 &=& -\frac{1}{2}\,z\,u_x-t\,u_z,
\\
\psi_5 &=& -u_z,
\\
\psi_6 &=& -u_x
\end{array}
\end{equation*}
where $A=A(t,y)$ and $B=B(t,y)$ below are arbitrary functions of $t$ and $y$. The commutators of
the generators are given by equations
\begin{equation*}
\{\phi_i(A), \phi_j(B)\} = \phi_{i+j}(A B_y-B A_y),
\end{equation*}
\begin{equation*}
\begin{array}{rclcrlc}
\{\psi_1, \phi_0(A)\} &=& \phi_0(t A_t),&&
\{\psi_1, \phi_1(A)\} &=& \phi_1(t A_t),
\\
\{\psi_2, \phi_0(A)\} &=& \phi_0(A_t),&&
\{\psi_2, \phi_1(A)\} &=& \phi_1(A_t),
\\
\{\psi_3, \phi_0(A)\} &=& 0,&&
\{\psi_3, \phi_1(A)\} &=& -\phi_1(A),
\\
\{\psi_4, \phi_0(A)\} &=& -\phi_1(t A_t),&&
\{\psi_4, \phi_1(A)\} &=& 0,
\\
\{\psi_5, \phi_0(A)\} &=& -\phi_1(A_t),&&
\{\psi_5, \phi_1(A)\} &=& 0,
\\
\{\psi_6, \phi_0(A)\} &=& 0,&&
\{\psi_6, \phi_1(A)\} &=& 0,
\end{array}
\end{equation*}

\begin{equation*}
\begin{array}{rclcrlc}
\{\psi_1, \psi_2\} &=& -\psi_2,&&
\{\psi_1, \psi_4\} &=& \psi_4,
\\
\{\psi_1, \psi_6\} &=& \psi_6,&&
\{\psi_2, \psi_4\} &=& \psi_5,
\\
\{\psi_3, \psi_4\} &=& -\psi_4,&&
\{\psi_3, \psi_5\} &=& -\psi_5,
\\
\{\psi_3, \psi_6\} &=& -2\,\psi_6,&&
\{\psi_4, \psi_5\} &=& -\frac{1}{2}\,\psi_5,
\end{array}
\end{equation*}
while $\{\psi_i, \psi_j\} = 0$ for all the other pairs $i < j$.
The commutator table shows that  the contact symmetry algebra of equation \eqref{4D_UHE}  is a semi-direct sum
$\mathfrak{sym}(\EuScript{E}_2) = \mathfrak{c}_\infty \rtimes \mathfrak{c}_\diamond$
of the six-dimensional Lie algebra
$\mathfrak{c}_\diamond =  \langle \psi_i \,\,\vert \,\, i \in \{1, \dots, 6\}  \rangle$
and the infinite-dimensional ideal
$\mathfrak{c}_\infty =
\EuScript{D}^{3}(\mathfrak{sym}(\EuScript{E}_2))
= \langle \phi_k(A) \,\,\vert \,\, k \in \{0, 1\}
\rangle$,
which, in its turn, is isomorphic to the tensor product $\mathbb{R}_2[h] \otimes \mathfrak{q}$
of the algebra of truncated polynomials $\mathbb{R}_2[h] =\mathbb{R}[h] / \langle h^2 = 0 \rangle$
and the Lie algebra $\mathfrak{q}$ of the vector fields of the form $A(t,y)\,\partial_y$.

\vskip 5 pt
\noindent
{\sc Remark 3.}
The spectral parameter $\lambda$ in the covering \eqref{Pavlov_Stoilov_covering} is non-removable. Indeed, the
symmetry $\psi_4$ of equation \eqref{4D_UHE} is unliftable to a symmetry of the covering
\eqref{Pavlov_Stoilov_covering}. For the associated vector field
$V(\psi_4) = \frac{1}{2}\,z\,\partial_x + t\,\partial_z$ the action of $\mathrm{exp}(2\,\lambda\,V(\psi_4))$ to
the Wahlquist--Estabrook form $\omega =  dv-u_z\,v_y\,dt -v_x\,dx - v_y\,dy - u_x\,v_y\,dz$ that defines
covering \eqref{Pavlov_Stoilov_covering} with $\lambda = 0$ gives
$e^{2\,\lambda V(\psi_4)} \omega=
dv-\left(\lambda^2\,v_x+(\lambda\,u_x+u_z)\,v_x\right)\,dt -v_x\,dx - v_y\,dy -(\lambda\,v_x+u_x\,v_y)\,dz$.
This is the Wahlquist--Estabrook form of covering \eqref{Pavlov_Stoilov_covering} with $\lambda \in \mathbb{R}$.
\hfill $\diamond$
\vskip 5 pt

Define the Maurer--Cartan forms $\beta_i$, $\theta_{k,m,n}$, $i \in \{1, \dots, 6\}$, $k \in \{0, 1\}$,
$m, n \in \mathbb{Z}_{+}$, of the Lie algebra  of $\mathfrak{sym}(\EuScript{E}_2)$ as dual forms to
its basis $\psi_i$, $\phi_k(t^m y^n)$, that is, put
$\beta_i(\psi_j) = \delta_{ij}$, $\beta_i(\phi_k(t^m y^n)) = 0$,
$\theta_{k,m,n}(\psi_i) = 0$, $\theta_{k,m,n}(\phi_p(t^q y^r)) = \delta_{kp} \delta_{mq} \delta_{nr}$.
Denote
$\Theta_k = \sum \limits_{m=0}^{\infty} \sum \limits_{n=0}^{\infty}
\frac{h_1^m}{m!} \frac{h_2^n}{n!} \theta_{k,m,n}$
for the formal parameters $h_1$, $h_2$ such that $dh_i=0$. Then the Maurer-Cartan structure equations of
$\mathfrak{sym}(\EuScript{E}_2)$ read
\begin{equation}
d\beta_1 = 0,
\label{4D_UHE_se_1}
\end{equation}
\begin{equation}
d\beta_2 = \beta_1 \wedge \beta_2,
\label{4D_UHE_se_2}
\end{equation}
\begin{equation}
d\beta_3 = 0,
\label{4D_UHE_se_3}
\end{equation}
\begin{equation}
d\beta_4 = (\beta_3-\beta_1) \wedge \beta_4,
\label{4D_UHE_se_4}
\end{equation}
\begin{equation}
d\beta_5 = \beta_3 \wedge \beta_5-\beta_2 \wedge \beta_4,
\label{4D_UHE_se_5}
\end{equation}
\begin{equation}
d\beta_6 = (2\,\beta_3- \beta_1) \wedge \beta_6+
\case{1}{2}\,\beta_4 \wedge \beta_5.
\label{4D_UHE_se_6}
\end{equation}
\begin{equation}
d\Theta_0 = \nabla_2(\Theta_0) \wedge \Theta_0 + \nabla_1(\Theta_0) \wedge (\beta_2 +h_1\,\beta_1)
\label{4D_UHE_se_7}
\end{equation}
\[
d\Theta_1 = \nabla_2(\Theta_1) \wedge \Theta_0+\nabla_2(\Theta_0) \wedge \Theta_1
+(\beta_1-\beta_3) \wedge \Theta_1
\]
\begin{equation}
\qquad\qquad
+(\beta_2+h_1\,\beta_1) \wedge \nabla_1(\Theta_1)
+(\beta_5+h_1\,\beta_4) \wedge \nabla_1(\Theta_0).
\label{4D_UHE_se_8}
\end{equation}

\subsection{Non-central extensions, Maurer--Cartan forms and Lax representation}

From the structure equations \eqref{4D_UHE_se_1} -- \eqref{4D_UHE_se_8} it follows that
$H^1(\mathfrak{sym}(\EuScript{E}_2)) = \langle \beta_1, \beta_3\rangle$ and
\begin{equation*}
H^2_{c_1\,\beta_1 + c_2 \,\beta_3}(\mathfrak{c}_\diamond) =
\left\{
\begin{array}{lcll}
\langle
[\beta_2 \wedge \beta_5]
\rangle,
&&
c_1 =1, &c_2 =1,
\\
&&
\\
\langle
[\beta_1 \wedge \beta_2],
[\beta_2 \wedge \beta_3]\rangle,
&~~~~&
c_1 =1,&c_2 =0,
\\
&&
\\
\langle
[\beta_1 \wedge \beta_4],
[\beta_3 \wedge \beta_4]
\rangle,
&&
c_1 =-1,&c_2 =1,
\\
&&
\\
\langle
[\beta_4 \wedge \beta_6]
\rangle,
&&
c_1 =-2,&c_2 =3,
\\
&&
\\
\{[0]\},
&&
\mathrm{otherwise}.&
\end{array}
\right.
\end{equation*}
Moreover, all the nontrivial exotic 2-cocycles of $\mathfrak{c}_\diamond$ are nontrivial exotic 2-cocycles
of $\mathfrak{sym}_0(\EuScript{E}_2)$ as well. Therefore they define a non-central extension
$\widehat{\mathfrak{c}}_\diamond$ of the Lie algebra $\mathfrak{c}_\diamond$ and hence a non-central extension
$\mathfrak{c}_\infty \rtimes \widehat{\mathfrak{c}}_\diamond$ of the Lie algebra
$\mathfrak{sym}_0(\EuScript{E}_2)$. The additional Maurer--Cartan forms $\beta_7$, ... , $\beta_{12}$ for the
extended Lie algebra are solutions to the system
\begin{equation}
d\beta_ 7 = (\beta_1+\beta_3) \wedge \beta_7 +\beta_2 \wedge \beta_5,
\label{4D_UHE_se_9}
\end{equation}
\begin{equation}
d\beta_8 = \beta_1 \wedge \beta_8 + \beta_1 \wedge \beta_2,
\label{4D_UHE_se_10}
\end{equation}
\begin{equation}
d\beta_9 = \beta_1 \wedge \beta_9 + \beta_2 \wedge \beta_3,
\label{4D_UHE_se_11}
\end{equation}
\begin{equation}
d\beta_{10} = (\beta_3-\beta_1) \wedge \beta_{10} + \beta_1 \wedge \beta_4,
\label{4D_UHE_se_12}
\end{equation}
\begin{equation}
d\beta_{11} = (\beta_3-\beta_1) \wedge \beta_{11} + \beta_3 \wedge \beta_4,
\label{4D_UHE_se_13}
\end{equation}
\begin{equation}
d\beta_{12} = (3\,\beta_3-2\,\beta_1) \wedge \beta_{12} + \beta_4 \wedge \beta_6.
\label{4D_UHE_se_14}
\end{equation}
This system is automatically compatible with equations \eqref{4D_UHE_se_1} -- \eqref{4D_UHE_se_8}.

Combining direct integration of the structure equations with Cartan's method of equivalence we  get the
explicit expressions for the Maurer--Cartan forms
\[
\beta_1 = \frac{dq}{q},
\quad
\beta_2 =q \,dt,
\quad
\beta_3 = \frac{dr}{r},
\quad
\beta_4 = \frac{r\,ds}{q},
\quad
\beta_5 = r\,(dz+2\,s\,dt),
\]
\[
\beta_6 = \frac{r^2}{q}\,(dx + s\,dz + s^2\,dt),
\quad
\theta_{0,0,0} = a_0\,(dy+a_1\,dt),
\]
\[
\theta_{1,0,0} = \frac{a_0\,r}{q}\,(du - a_1\,dz - a_2\,dy-a_3\,dt),
\quad
\beta_{10} = \frac{r}{q} \,(d a_4 +\ln q \,ds),
\]
as well as the requirement for the linear combination
\[
\theta_{1,0,0} -\beta_6 =  \frac{a_0\,r}{q}\,\left(du-\frac{r}{a_0}\,dx - a_2\,dy- \left(a_1+\frac{rs}{a_0}\right)\,dz-\left(a_3+ \frac{rs^2}{a_0}\right)\,dt\right)
\]
to be a multiple of the contact form $du-u_t\,dt - u_x\,dx-u_y\,dy-u_z\,dz$.
This gives the change of notation
$r= a_0\,u_x$,
$a_1 = u_z-s\,u_x$,
$a_2 = u_y$,
$a_3 = u_t - s^2\,u_x$,
which yields
\[
\theta_{1,0,0} -\beta_6 =  \frac{a_0^2\,u_x}{q}\,(du-u_t\,dt - u_x\,dx-u_y\,dy-u_z\,dz).
\]

Our attempts to find a linear combination of the Maurer--Cartan forms $\beta_1$, ... , $\beta_{12}$,
$\theta_{k,m,n}$ have not given a Wahlquist--Estabrook form of any covering over equation \eqref{4D_UHE}.
Therefore we have extended the Lie algebra $\widehat{\mathfrak{c}}_\diamond$ with the structure equations
\eqref{4D_UHE_se_1} --  \eqref{4D_UHE_se_14} via the same procedure,
that is,  by finding nontrivial cocycles from
$H^2_{c_1 \beta_1 + c_2 \beta_3}(\widehat{\mathfrak{c}}_\diamond)$.
The additional Maurer--Cartan forms $\beta_{13}$, ... , $\beta_{30}$ of the resulting 18-dimensional non-central
extension of $\widehat{\mathfrak{c}}_\diamond$ and of
$\mathfrak{c}_\infty \rtimes \widehat{\mathfrak{c}}_\diamond$
are solutions to the structure equations
\begin{equation}
d\beta_{13}=2 \, (\beta_3-\beta_1) \wedge \beta_{13}+\beta_{4} \wedge \beta_{10},
\label{4D_UHE_se_15}
\end{equation}
\[
d\beta_{14}=(2 \, \beta_1+\beta_3) \wedge \beta_{14}+\beta_2 \wedge \beta_{7},
\]
\[
d\beta_{15}=2 \, \beta_1 \wedge \beta_{15}+\beta_2 \wedge \beta_{8},
\]
\[
d\beta_{16}=2 \, \beta_1 \wedge \beta_{16}+\beta_2 \wedge \beta_{9},
\]
\[
d\beta_{17}=\beta_1 \wedge \beta_{17}+\beta_1 \wedge \beta_{8},
\]
\[
d\beta_{18}=\beta_1 \wedge \beta_{18}+\beta_1 \wedge \beta_{9}+\beta_3 \wedge \beta_{8},
\]
\[
d\beta_{19}=\beta_1 \wedge \beta_{19}+\beta_3 \wedge \beta_{9},
\]
\[
d\beta_{20}=2 \, \beta_3 \wedge \beta_{20}+2\,\beta_2 \wedge \beta_{6}+\beta_{4} \wedge \beta_{7},
\]
\[
d\beta_{21}=\beta_3 \wedge \beta_{21}+\beta_1 \wedge \beta_{5}+\beta_{4} \wedge \beta_{8},
\]
\[
d\beta_{22}=\beta_3 \wedge \beta_{22}+\beta_2 \wedge \beta_{10}+\beta_{4} \wedge \beta_{8},
\]
\[
d\beta_{23}=\beta_3 \wedge \beta_{23}+\beta_2 \wedge \beta_{11}+\beta_{4} \wedge \beta_{9},
\]
\[
d\beta_{24}=\beta_3 \wedge \beta_{24}+\beta_3 \wedge \beta_{5}+\beta_{4} \wedge \beta_{9},
\]
\[
d\beta_{25}=(3 \, \beta_3 -\beta_1) \wedge \beta_{25}+\beta_2 \wedge \beta_{12}-\beta_{5} \wedge \beta_{6},
\]
\[
d\beta_{26}=(\beta_3-\beta_1) \wedge \beta_{26}+\beta_1 \wedge \beta_{11}+\beta_3 \wedge \beta_{10},
\]
\[
d\beta_{27}=(\beta_3-\beta_1) \wedge \beta_{27}+\beta_3 \wedge \beta_{11},
\]
\[
d\beta_{28}=(\beta_3-\beta_1) \wedge \beta_{28}+\beta_1 \wedge \beta_{10},
\]
\[
d\beta_{29}=2 \, (\beta_3-\beta_1) \wedge \beta_{29}+\beta_{4} \wedge \beta_{11},
\]
\[
d\beta_{30}=(4 \, \beta_3 -3 \, \beta_1) \wedge \beta_{30}+\beta_{4} \wedge \beta_{12}.
\]
These equations are compatible with equations \eqref{4D_UHE_se_1} -- \eqref{4D_UHE_se_14}.
Integration of equation \eqref{4D_UHE_se_15} gives
$\beta_{13} =a_0^2\,u_x^2 q^{-2}\,(dv-a_4\,ds)$.
Then the linear combination
\[
\fl
\beta_{13} - \theta_{0,0,0} -\beta_5 - \beta_6 =
\]
\[
\fl
\quad\quad
\frac{a_0^2u_x^2}{q^2}\,\left(dv - a_4\,ds -q\,dx -\frac{q^2}{a_0u_x^2}\,dy
-\frac{q+a_0\,s\,u_x}{a_0u_x}\,dz
-\frac{q\,(q\,u_z+s\,q\,u_x+a_0\,s^2\,u_x^2)}{a_0u_x^2}\,dt
\right)
\]
after the change of notation
$q=v_x$,
$a_0 = v_x^2 u_x^{-2}v_y^{-1}$,
$a_4 = v_s$
gives
\[
\fl
\beta_{13} - \theta_{0,0,0} -\beta_5 - \beta_6 =
\]
\[
\fl
\quad\quad
\frac{v_x^2}{v_y^2u_x^2}\,\left(dv - v_s\,ds -v_x\,dx -v_y\,dy
-(s\,v_x+u_x\,v_y)\,dz
-(s^2\,v_x+(u_z+s\,u_x)\,dt
\right),
\]
which is the Wahlquist--Estabrook form of the covering
\[
\left\{
\begin{array}{lcl}
v_t&=& s^2\,v_x +(s\,u_x+u_z)\,v_y,
\\
v_z&=& s\,v_x+u_x\,v_y.
\end{array}
\right.
\]
This system differs from \eqref{Pavlov_Stoilov_covering} by the notation.

\vskip 10 pt
\noindent
{\sc Remark 4}.
The non-removable parameter $s$ in the above system has appeared during com\-pu\-tation of
 form $\beta_4$, which is dual to the unliftable symmetry $\psi_4$, cf. Remark 3.
\hfill $\diamond$
\vskip 5 pt

\section{Conclusion}
In the present paper we have shown that the method of \cite{Morozov2017,Morozov2018a} is applicable to
Lax re\-pre\-sen\-ta\-ti\-ons with non-removable parameters, in particular, the Lax representations for
equations \eqref{Pavlov_eq} and \eqref{4D_UHE} can be derived from  the non-central extensions of contact
symmetry algebras of these equations. In both examples the symmetry al\-ge\-b\-ras have the specific
structure of the semi-direct product of an infinite-dimensional ideal and a non-Abelian
finite-di\-men\-si\-o\-nal Lie subalgebra. The cohomological properties of the finite-dimensional subalgebras
appear to be sufficient to reveal non-central extensions that define the Lax representations.

It is natural to ask whether the method can produce the known as well as new Lax representations for other
equations. Also, we expect that the proposed technique will be helpful in describing multi-component
integrable generalizations of integrable {\sc pde}s,
\cite{Dunajski2002,ManakovSantini2006,Bogdanov2010,Morozov2012,KruglikovMorozov2015}.
We intend to address these issues in the further study.

\section*{Acknowledgments}

This work was partially supported by the Faculty of Applied Mathematics of AGH UST statutory tasks within
subsidy of Ministry of Science and Higher Education.

I am very grateful to I.S. Krasil${}^{\prime}$shchik for useful discussions.
I thank L.V. Bog\-da\-nov  and P. Zusmanovich for important remarks.

\end{document}